\documentclass[12pt,letterpaper]{article}

\usepackage{amsmath,amssymb}
\usepackage{graphicx}
\usepackage[utf8]{inputenc}
\usepackage{placeins}
\usepackage{caption}
\usepackage{subcaption}
\usepackage{geometry}

\usepackage[style=numeric-comp,sorting=none]{biblatex}
\usepackage{amsmath,amssymb}
\DeclareMathOperator{\E}{\mathbb{E}}
\newcommand{\Nor}{\mathcal{N}}
\newcommand{\Corr}{{\rm Corr}}
\newcommand{\Cov}{{\rm Cov}}
\newcommand{\Var}{{\rm Var}}
\newcommand{\bs}{{\backslash}}
\newcommand{\Cbf}{{\mathbf C}}
\newcommand{\rhohat}{{\widehat\rho}}

\title{Optimal Cross-Correlation Estimates from Asynchronous Tick-by-Tick Trading Data}
\author{William H. Press\\
Oden Institute for Computational Engineering and Sciences\\
The University of Texas at Austin}

\begin{document}
\maketitle

\begin{abstract}
Given two time series, $A$ and $B$, sampled asynchronously at different times $\{t_{Ai}\}$ and $\{t_{Bj}\}$, termed ``ticks", how can one best estimate the correlation coefficient $\rho$ between changes in $A$ and $B$? We derive a natural, minimum-variance estimator that does not use any interpolation or binning, then derive from it a fast (linear time) estimator that is demonstratably nearly as good. This ``fast tickwise estimator" is compared in simulation to the usual method of interpolating changes to a regular grid. Even when the grid spacing is optimized for the particular parameters (not often possible in practice), the fast tickwise estimator has generally smaller estimation errors, often by a large factor. These results are directly applicable to tick-by-tick price data of financial assets.
\end{abstract}

\section{Introduction}

Correlations among the the market price movements of financial assets play a central role in modern finance in areas as diverse as portfolio construction, high-frequency trading strategies, and options pricing. If instantaneous ``true" prices of multiple assets were available at arbitrary times $t$, then standard statistical methods for uniformly time-sampled series could be used to estimate correlations from empirical data. Such methods are well studied.\cite{IntroBook,Jong}

However, market asset prices are (by definition) known only when trades occur, a set of times $\{t_i\}$ that is different for each asset, termed asynchronous. The study of the effects of asynchronicity on understanding markets has a long history. A landmark is the 1990 paper of Lo and MacKinlay\cite{LoMac}, cited more than a thousand times. Still, the specific issue of how best to estimate the pairwise correlation between two assets trading asynchronously is not completely settled, as evidenced by the continuing flow of papers on the subject and its variations. Correlation estimation methods studied include various summed cross-products \cite{Jong,RealCorr,Ole}, interpolation or extension to a grid of common times \cite{IntroBook,Bonanno}, other binned and kernel methods \cite{Geo}, most-recent-tick intervals \cite{Zhang}, Fourier transform methods \cite{Reno,Precup2,Precup1}, and others---this taxonomy neither exhaustive nor mutually exclusive.  
This paper, in the general class of summed cross-products, attempts to add clarity by showing that, under simple model assumptions, there is a simple, natural best estimator in a specified sense that we will define.

\subsection{Correlated, Memoryless Gaussian Process Model}

We model the returns of two assets $A$ and $B$ by the widely-studied model of memoryless, correlated, continuous-time Gaussian processes of zero mean and unit variance per unit time (possibly thus after detrending and rescaling as discussed below in \S\ref{sec34}). We adopt the symbol $\Nor$ as denoting a draw from the normal distribution $N(0,1)$ and further adopt the convention that different subscripts $\{\Nor_x,\Nor_y\}$ represent independent draws, identical subscripts $\{\Nor_z,\Nor_z\}$ denote the same draw (that is, have the same numerical value), and the special notation $\Nor_\sharp$ denotes draws that are each independent of all other draws in a given set of expressions (both $\Nor_\sharp$'s and conventionally labeled $\Nor$'s). With this notation,
\begin{equation}
    \E(\Nor_a) = 0, \quad \E(\Nor_a\Nor_b) = 0, \quad \E(\Nor_c^2) = 1, , \quad \E(\Nor_\sharp\Nor_\sharp) = 0, \quad \E(\Nor_c^4) = 3
\label{norrules}
\end{equation}
where only the last is not trivial (but is well known, e.g. from Isserlis' theorem).

The  respective total returns for $A$ and $B$ over a specified time interval $(t_1,t_2)$ can now be written as ($\sim$ meaning ``drawn from"),
\begin{equation}
\begin{split}
    R_A &\sim \sqrt{\Delta}\,\left( \sqrt{\rho} \Nor_{c} + \sqrt{1-\rho} \,\Nor_\sharp   \right) \\
    R_B &\sim \sqrt{\Delta}\,\left( \sqrt{\rho} \Nor_{c} + \sqrt{1-\rho} \,\Nor_\sharp   \right)   
\label{eq1}
\end{split}
\end{equation}
Here $\Delta \equiv t_2-t_1$, and $\rho = \Corr(R_A,R_B)$ is the specified correlation (here without loss of generality taken as positive for notational convenience).   One easily sees that the normalized product
\begin{equation}
    \frac{R_A R_B}{\Delta} \equiv \rhohat 
\label{eq3}
\end{equation}
is an unbiased estimator of $\rho$ in the interval $(t_1,t_2)$, since
\begin{equation}
    \E\left(\frac{R_A R_B}{\Delta}\right) = \E(\rho \Nor_c^2) = \rho \E(\Nor_c^2) = \rho
\end{equation}
But equation \eqref{eq3} applies only in the uninteresting synchonous case that $t_1$ and $t_2$ are trading ticks for both $A$ and $B$.

The remaining plan of this paper can now be stated:
\begin{itemize}
\item Generalize equation \eqref{eq3} to the case of non-identical intervals for the measured returns $R_A,R_B$. This will yield a plethora of unbiased estimators $\rhohat$, in principle one for every pairing of an A market tick with a B market tick.
\item Argue that a ``natural" unbiased estimator for $\rho$ must be some weighted mean of all of these individual estimators, because that class includes {\em all} bilinear expressions in the returns of $A$ and returns of $B$ over {\em any} multiple-tick intervals of each.
\item Then, derive a formula for those weights (for the weighted mean) that give a minimum-variance estimator, i.e., that minimize $\Var(\rhohat)$. A surprisingly simple result is obtained.
\item  For the case that the tick times for $A$, $\{t_{Ai}\}$, and for $B$, $\{t_{Bj}\}$ are independently Poisson random (with possibly different rates), we obtain from simulations further simple, heuristic rules for some optimal and near-optimal correlation estimates, including one computable in linear time. 
\end{itemize}

\begin{figure}[ht]
\centering
\includegraphics[width=12cm]{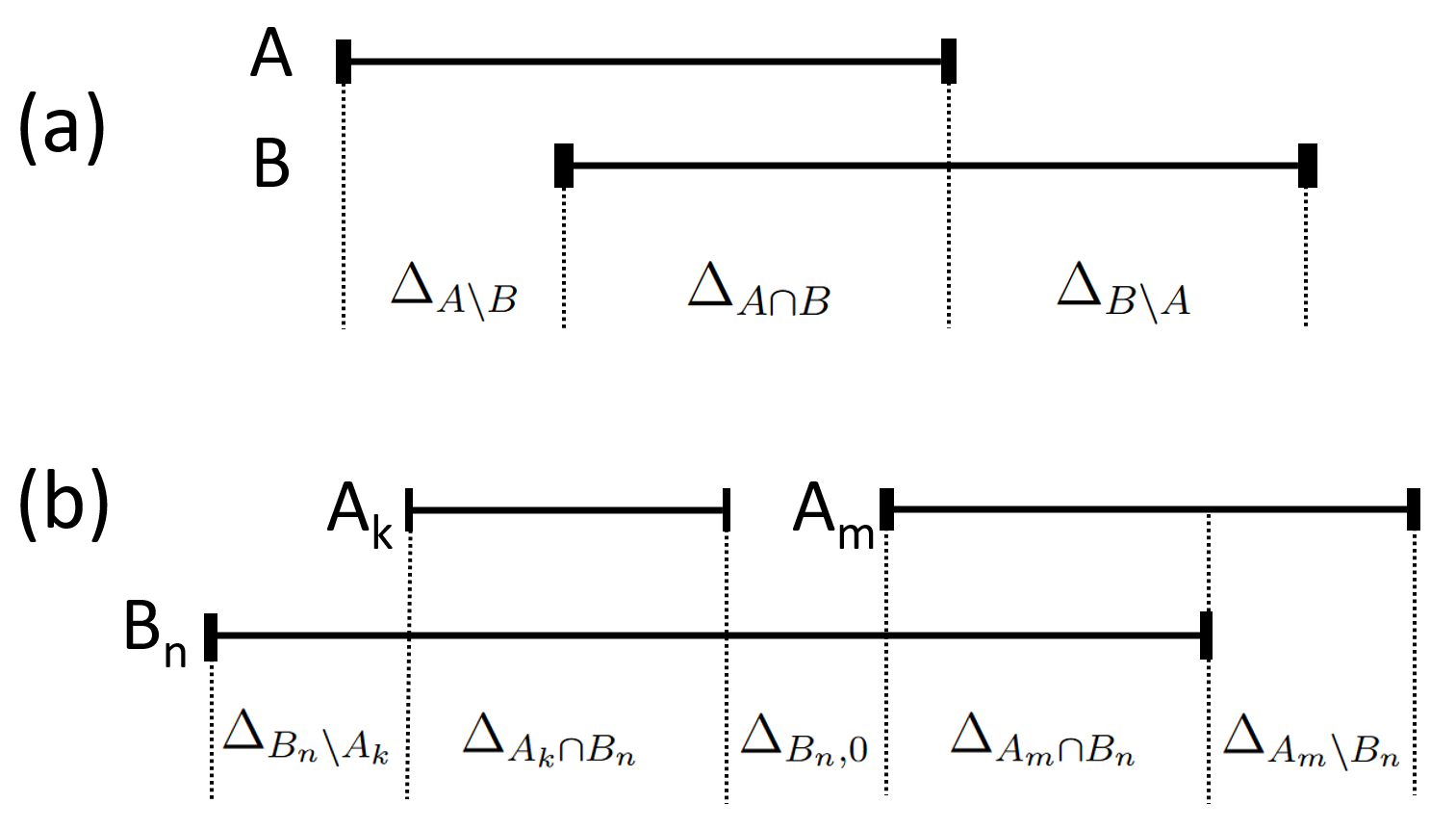}
\caption{(a) Diagram showing non-overlapping and overlapping time intervals for partially overlapping $A$ and $B$ tick intervals, notation used in equation \eqref{eq5}. (b) Time intervals when two or more $A$ segments overlap a single $B$ segment, notation used in equation \eqref{eq17}. In the configuration drawn, the two other possible intervals $\Delta_{A_k\bs B_n}$ and $\Delta_{B_n\bs A_m}$ don't exist, i.e., have zero length.}
\label{fig0}
\end{figure}

\section{Preliminaries}
\subsection{Expectation and Variance of Cross-Products of Returns}

We here calculate $\E(R_AR_B)$ and $\Var(R_AR_B)$, where $R_A$ is the return of $A$ over the interval $(t_1,t_2)$, $R_B$ the return of $B$ over the interval $(t_3,t_4)$. Define the interval lengths $\Delta_A \equiv t_2-t_1$ and $\Delta_B \equiv t_4-t_3$. We
can immediately dispose of the vast majority of all such cross-products, namely those with non-overlapping time segments. By the assumption of memorylessness for $A$ and $B$, these $R_A$'s and $R_B$'s are independent normals $\Nor\sharp$ yielding no information about $\rho$, $\E(R_AR_B) = 0$.

That leaves as interesting cases those where the time segments overlap. With $t_1 \le t_3 < t_2 \le t_4$, the overlap interval is $(t_3,t_2)$ whose length is $\Delta_{A\cap B} \equiv t_2-t_3$. With the obvious notation $\Delta_{A\bs B} = t_3-t_1$ and $\Delta_{B\bs A} = t_4-t_2$ (see Figure \ref{fig0}(a)), the analog of equation \eqref{eq1} is
\begin{equation}
\begin{split}
    R_A &\sim \sqrt{\Delta_{A\bs B}}\,\Nor_\sharp + \sqrt{\Delta_{A\cap B}}\,\left( \sqrt{\rho} \Nor_{c} + \sqrt{1-\rho} \,\Nor_\sharp \right) \\
    R_B &\sim \sqrt{\Delta_{A\cap B}}\,\left( \sqrt{\rho} \Nor_{c} + \sqrt{1-\rho} \,\Nor_\sharp   \right) + \sqrt{\Delta_{B\bs A}}\,\Nor_\sharp
\label{eq5}
\end{split}
\end{equation}

Using equation \eqref{norrules}, one calculates straightforwardly the surprisingly simple results,
\begin{subequations}
\begin{align}
\E(R_AR_B) &= \Delta_{A\cap B}\, \rho \label{eq6a}\\
\Var(R_AR_B) &= \E(R_A^2R_B^2) - \E(R_AR_b)^2 = \Delta_A\Delta_B + \Delta_{A\cap B}^2 \,\rho\label{eq6b}
\end{align}
\label{eq6}
\end{subequations}
whence the estimator (and its variance),
\begin{subequations}
\begin{align}
&\rhohat = \frac{R_AR_B}{\Delta_{A\cap B}}\label{eq7a}\\
&\Var(\rhohat) = \frac{\Delta_A\Delta_B}{\Delta_{A\cap B}^2} + \rho^2\label{eq7b}
\end{align}
\label{eq7}
\end{subequations}
Notice that equation \eqref{eq7a} essentially recapitulates equation \eqref{eq3} for $\rhohat$, but with the denominator now the length of only the segments' overlap. Every pair of overlapping segments thus gives a separate (not necessarily independent, see next section) estimator for $\rhohat$. The variance of that estimator is large if the overlap length is small compared with either segment. In equation \eqref{eq7b} the term $\rho^2$ is never large compared to the first term, especially when dealing with small correlations. Also, note that $\rho^2$ is a population value, not known a priori. With many pairs of overlapping segments producing a weighted average $\widehat\rhohat$ (to be defined below), we will propose (and will validate in simulation) evaluating equation \eqref{eq7b} iteratively, starting with $\rho^2 = 0$.

\subsection{Variance of Arbitrarily Weighted Means}

We here digress to recall, for use below, some standard results, all easy to prove.

For positive weights $w_i$, $i=1,\ldots,N$, let $Q$ be the weighted mean
\begin{equation}
    Q = \sum_{i=1}^N w_i\,q_i \Biggm/ \sum_{i=1}^N w_i
\label{eq8}
\end{equation}
Then for the variance of the weighted mean, we have
\begin{equation}
\begin{split}
\Var(Q) &= \E(Q^2) - \E(Q)^2 \\
&= \left[ \sum_{ij} w_iw_j\E(q_iq_j)
- \left( \sum_i \E(q_i)\right)^2\right] \Biggm/ \left( \sum_i w_i \right)^{2}\\
&= \sum_{ij} w_iw_j \Cov(q_i,q_j) \Biggm/ \left( \sum_i w_i \right)^{2}
\end{split}
\label{eq9}
\end{equation}
where
\begin{equation}
    \Cov(q_i,q_j) \equiv C_{ij} = \E(q_iq_j) - \E(q_i)E(q_j)
\end{equation}
is the $i,j$ component of the covariance matrix of the $q$'s.
In the elementary case $w_i = \text{const}$, equation \eqref{eq9} reduces to
\begin{equation}
    \Var(Q) = \frac{1}{N^2}\sum_{ij} C_{ij}
\end{equation}
that is, the mean of all the $C_{ij}$'s, denoted $\left< C_{ij} \right>$, If $\Cbf$ is diagonal, then this reduces further to $\left< C_{ii} \right>/N$, recovering the familiar rule that the variance of the unweighted mean of $N$ independent quantities is their mean variance divided by $N$.

One is frequently advised to use weights proportional to the inverse variance, $w_i = 1/C_{ii}$. Then, equation \eqref{eq9} yields
\begin{equation}
   \Var(Q) = \frac{\sum_{ij} C_{ij}/(C_{ii}C_{jj})}{\left[\sum_i (1/C_{ii})\right]^2}
\label{varwgt}
\end{equation}
Notice (relevant below) that this expression does not require calculating the inverse matrix $\Cbf^{-1}$.

Finally, let us derive what are the optimal weights, those that minimize $\Var(Q)$. Without loss of generality~let $\sum_i w_i=1$. Then, with Lagrange multiplier $2 \lambda$, from equation \eqref{eq9}, we minimize the Lagrangian
\begin{equation}
    \mathcal{L} = \sum_{ij} w_i C_{ij} w_j - 2\lambda \sum_i w_i
\end{equation}
w.r.t. $w_i$, giving
\begin{equation}
    \sum_{j} C_{ij} w_j = \lambda
\end{equation}
Left-multiplying by the inverse matrix $\Cbf^{-1}$ gives, with equation \eqref{eq9},
\begin{equation}
    w_j = \sum_i [\Cbf^{-1}]_{ij},\quad \text{and} \quad
    \Var(Q) = 1 \Big/ \sum_{ij} [\Cbf^{-1}]_{ij}
\label{optwgt}
\end{equation}
One easily sees that equation \eqref{optwgt} reduces to equation \eqref{varwgt} if the covariance $\Cbf$ is diagonal, i.e., if
the $q_i$'s are all independent. This shows that inverse-variance weighting is optimal for that case, the well-known result.

Now back to our estimators $\rhohat$.

\subsection{Covariance of the Overlapping-Segment Estimators}
\label{sec23}

Let $I=1,2,\ldots,M$ index all distinct pairs $P_I$ of overlapping $A$ and $B$ intervals. In terms of tick times, one value of $I$, for example, might denote the pair $[(t^A_{i},t^A_{i+1}),$ $(t^B_{j},t^B_{j+1}))]$, if $t^B_{j} < t^A_{i+1}$. The notation $P_I = [A_i,B_j]$ will be useful. Up to edge effects of $\pm 1$, the number of distinct pairs $M$ equals the cardinality of the combined set $\#\{\{t^A_i\}\cup\{t^B_j\}\}$, because the distinct start of each $A$ or $B$ segment begins a new distinct pair. (In fact, this observation leads to an efficient algorithm for finding all the pairs in large data sets of ticks.)

Let $\rhohat_I$ denote the estimator from equation \eqref{eq7a} using segment pair $I$. We have equation \eqref{eq7b} for the variance $\Var(\rhohat_I)$, but what about the covariance $\Cov(\rhohat_I,\rhohat_J)$, $I \neq J$?

Let $P_I = [A_k,B_l]$, $P_J = [A_m,B_n]$. Then, if $k\ne m$ and $l\ne n$, the estimators share no segment in either $A$ or $B$, hence are independent with zero covariance. On the other hand, if $k = m$ and $l = n$, then $I = J$, and we already know the variance. So the only interesting case is $k \ne m$ and $l = n$ (or the equivalent case exchanging $A$ and $B$). A list of all possible overlap lengths (including some that, geometrically, must be zero if others are nonzero) is, extending the previous notation in the obvious way,
\begin{equation}
    \Delta_{A_k\bs B_n}, \Delta_{B_n\bs A_k},\Delta_{A_k\cap B_n},\Delta_{B_n,0},\Delta_{A_m\cap B_n},\Delta_{A_m\bs B_n},\Delta_{B_n\bs A_m}
\end{equation}
where $\Delta_{B_n,0}$ denotes the (possibly zero) length within $B_n$ and between $A_k$ and $A_m$. Figure \ref{fig0}(b) shows how these different segments fit together.

The analog of equation \eqref{eq5} is now
\begin{equation}
\begin{split}
    R_{A_k} &\sim \sqrt{\Delta_{A_k\bs B_n}}\,\Nor_\sharp + \sqrt{\Delta_{A_k\cap B_n}}\,\left( \sqrt{\rho} \Nor_{k} + \sqrt{1-\rho} \,\Nor_\sharp \right) \\
    R_{A_m} &\sim \sqrt{\Delta_{A_m\cap B_n}}\,\left( \sqrt{\rho} \Nor_{m} + \sqrt{1-\rho} \,\Nor_\sharp \right) + \sqrt{\Delta_{A_m\bs B_n}}\Nor_\sharp\\
    R_{B_n} &\sim \sqrt{\Delta_{B_n\bs A_k}}\Nor_\sharp + \sqrt{\Delta_{A_k\cap B_n}}\,\left( \sqrt{\rho} \Nor_{k} + \sqrt{1-\rho} \,\Nor_\sharp \right) + \sqrt{\Delta_{B_n,0}} \Nor_\sharp\\
    &\qquad + \sqrt{\Delta_{A_m\cap B_n}}\,\left( \sqrt{\rho} \Nor_{m} + \sqrt{1-\rho} \,\Nor_\sharp \right) + \sqrt{\Delta_{B_n\bs A_m}}\Nor_\sharp 
\label{eq17}
\end{split}
\end{equation}
Straightforward calculation gives
\begin{equation}
\begin{split}
\E(A_k A_m) &= 0 \quad \text{(non-overlapping $A$ segments)}\\
\E(A_k B_n) &= \Delta_{A_k\cap B_n} \rho \quad\text{(as from equation \eqref{eq6a})}\\
\E(A_m B_n) &= \Delta_{A_m\cap B_n} \rho \quad\text{(ditto)}\\
\Cov(A_k B_n,A_m B_n) &= \E(A_k B_n^2 A_m)- \E(A_k B_n)\E(A_m B_n)\\
&= \Delta_{A_k\cap B_n}\Delta_{A_m\cap B_n}\rho^2
\end{split}
\end{equation}
implying
\begin{equation}
    \Cov(\rhohat_I,\rhohat_J) = \rho^2
\label{eq19}
\end{equation}
The remarkably simple result is that the covariance between estimators $\rhohat$ (equation \eqref{eq7a}) that share either an $A$ or a $B$ segment is just $\rho^2$, independent of any of the lengths of segments or their overlaps, and zero for any estimators that don't share any segment.

\section{Optimal and Fast Correlation Estimators}
\label{sec6}
Here we assemble the pieces developed, also mentioning the implied computational workload.

We are given an ordered series of time ticks for asset $A$, say $t^A_i$, $i=1,2,\ldots,M$ and for asset $B$, $t^B_j$, $j=1,2,\ldots,K$. Ties between $t^A_i$'s and $t^B_j$'s are allowed. The length of segment $i$ in $A$ or $B$ is $\Delta_{\{A,B\},i} \equiv t^{\{A,B\}}_{i+1} - t^{\{A,B\}}_i$. (We can suppress the $i$ index when its value is unambiguous, below.)

Without loss of generality assume $M \ge K$ (i.e., $A$ trades more frequently on average than $B$). Denote $A$'s total return in the interval $(t^A_i,t^A_{i+1})$ by $R^A_i$, and similarly for $R^B_j$. As previously, let $\{P_I\}$, $I=1,2,\ldots,N$ be the set of overlapping $A$ and $B$ segment pairs, $P_I = [A_i,B_j]$ for some $i(I)$ and $j(I)$. As already mentioned, $N \approx M+K$ (the approximation from edge effects of $\pm 1$). Finding the $P_I$'s is a calculation of linear order $O(M+K)$. Denote by $\Delta_{A\cap B,I}$ the length of overlap between the $A$ and $B$ segments in $P_I$. 

We now define two estimators for the correlation of assets $A$ and $B$, $\rho$, one optimal, the other fast. For each, we also give its expected variance. It will turn out that the fast estimator is almost as good as the optimal one. The variance is important to know, not least to manage the tradeoff between the locality of an estimate $\rhohat$ (which is better with fewer segments) and its accuracy (which is better with more segments). Knowing the variance, the user can make an appropriate choice. Also, these variances will demonstrate that our fast estimator is indeed nearly as good as the optimal one.

\subsection{Optimal Tickwise Estimator}
\label{sub31}
 The optimal (i.e., minimum variance) estimator follows from equation \eqref{optwgt}. The correlation matrix $C_{IJ}$ has diagonal elements
 \begin{equation}
     C_{II} = \frac{\Delta_{A,I} \Delta_{B,I}}{(\Delta_{A\cap B,I})^2} + \rho^2
\label{CII}
 \end{equation}
and off-diagonal elements
 \begin{equation}
     C_{IJ} = \begin{cases}
  \rho^2  & \text{if $I \ne J$ share an $A$ or $B$ segment}  \\
  0 & \text{otherwise}
\end{cases}
\label{CIJ}
 \end{equation}
 In general $\Cbf$ is band-diagonal. The bandwidth depends on the specific tick times, but will generally be of order $O(M/K)>1$, the ratio of trading rates, because about that many $B$ segments will overlap each $A$ segment.
Inverting the matrix $\Cbf$ thus requires $O[(M/K)(M+K)^2]$ operations and $O[(M+K)^2]$ storage. This may or may not be prohibitive, but will not be, in any case, what we ultimately will recommend.

The optimal, i.e., minimum variance, estimator using all segments is (equation \eqref{optwgt}),
\begin{equation}
\widehat\rhohat = \sum_I w_I \rhohat_I,\quad w_I = \sum_J [\Cbf^{-1}]_{IJ}
\label{rhoopt}
\end{equation}
with variance
\begin{equation}
\Var\left(\widehat\rhohat\right) = 1 \Bigm/ \sum_{I} w_I
\label{varopt}
\end{equation}
Here and below, the double-hatted $\widehat\rhohat$ denotes a single global estimate combining all available time segments---i.e., the desired answer.
Since the population value $\rho$ that enters equations \eqref{CII} and \eqref{CIJ} is not known a priori, a computational strategy is to compute equations \eqref{CII}--\eqref{rhoopt} with $\rho = 0$,
then iteratively recompute those equations with the obtained value $\rho = \widehat\rhohat$. Empirically, we find that this converges rapidly, in $\lesssim 3$ iterations.

\subsection{Fast Tickwise Estimator}
\label{sub32}
The expensive matrix inversion of $\Cbf$ is avoided if we can ignore its off-diagonal elements in calculating $\widehat\rhohat$, equivalent to employing the inverse-variance weighted mean of the $\rhohat_I$'s, a linear-time calculation $O(M+K)$,
\begin{equation}
\widehat\rhohat = \sum_I w_I \rhohat_I,\quad w_I = 1/C_{II}
\label{rhonearopt}
\end{equation}
The variance of this estimate is calculable without approximation (equation \eqref{varwgt}) as
\begin{equation}
\Var\left(\widehat\rhohat\right) = \frac{1}{\sum_I w_I} + 
\rho^2\, \frac{\sum_{\cal O} w_I w_J}{(\sum_I w_I)^2} 
\label{varnearopt}
\end{equation}
Here the index set ${\cal O}$ is defined by
\begin{equation}
    {\cal O} = \{ (I,J) \text{ s.t. } I\ne J 
    \text{ and $I$,$J$ share an $A$ or $B$ segment}  \}
\end{equation}
that is, the set of off-diagonal $(I,J)$'s populated by the values $\rho$.
Equation \eqref{varnearopt} is then calculable in $O[(M/K)(M+K)]$ operations. Below we will compare in simulation equations \eqref{varnearopt} and \eqref{varopt} and verify that they give nearly equal values.

\subsection{Some Special Cases}

\subsubsection*{Cases of Most Synchronous and Most Asynchronous}

Some intuition on the preceding results can be gained by considering two unrealistic cases: (i) $N$ exactly aligned $A$ and $B$ segments , and (ii) $N$ segments of $A$, all of equal length, with $B$'s $N-1$ segments beginning and ending at consecutive midpoints of the $A$'s.

In the first case, the $N$ estimates $\rhohat_I$ each have variance $1+\rho^2$ (equation \eqref{eq7b}) and the covariances are all zero. Thus, the optimal and fast estimates $\widehat\rhohat$ are identical, with variance
\begin{equation}
    \Var\left(\widehat\rhohat\right) = \frac{1+\rho^2}{N}
\label{eq27}
\end{equation}

In the second case, we have $2N$ estimates $\rhohat_I$, each with variance
\begin{equation}
    \Var\left(\rhohat_I\right) = \frac{L\cdot L}{(L/2)^2} + \rho^2 = 4 + \rho^2
\label{eq28}
\end{equation}
The correlation matrix $\Cbf$ is tridiagonal of size $2N\times 2N$, with diagonal elements $4+\rho^2$ and sub- and super-diagonal elements $\rho^2$.

We can use a trick to jump to the answer for large $N$: Alter the two corner elements of $\Cbf$, making $C_{1,2N} = C_{2N,1} = \rho^2$. (This is equivalent to wrapping time around in a circle.) The new matrix is a circulant with all rows and all columns now equivalent. By symmetry the weights $w_I$, including the optimal weights, must be constant. But equations \eqref{eq8} and \eqref{eq9} are invariant under scaling all the weights to unity, under which equation \eqref{eq9} gives immediately 
\begin{equation}
 \Var\left(\widehat\rhohat\right) = \frac{(2N)(4+\rho^2) + (4N-2)\rho^2}{(2N)^2} \approx \frac{2 + \tfrac{3}{2}\rho^2}{N}
\label{eq29}
\end{equation}
The above argument shows that the fast estimator must coincide with the optimal estimator in this highly symmetric case. Its (formally) different constant weights will also yield equation \eqref{eq29}.

Comparing equations \eqref{eq27} and \eqref{eq29}, one sees that the variance in the uniform asynchronous case is larger than the synchronous case by a factor decreasing from $2$ at $\rho^2=0$ to $7/4$ at $\rho=1$, Below we will give the analogous comparisons for randomly unstructured tick times.

\subsubsection*{Case with Large Difference in Trading Rates}

A different illuminating special case is to have many, say $M \gg 1$, tick intervals of $A$ comprising one $B$ interval. In the unrealistic case that the short $A$ intervals are all equal in length, then the $M\times M$ covariance matrix has diagonals $M+\rho^2$, off-diagonals all $\rho$. This is (without modification) a circulant, so the argument of the previous subsection applies. Equation \eqref{eq9} then trivially gives the result $\Var\left(\widehat\rhohat\right) = 1 + \rho^2$, which would also be the result obtained combining all the $A$ segments into a single segment congruent to the single $B$ interval. So, no information is lost either way.

\subsection{Self-Scaling Fast Correlation Estimator}
\label{sec34}

Up to now we have assumed that returns $R_A$ and $R_B$ are perfectly scaled to have unit variance per unit time. That allowed the previous exploration of unbiased estimators with (provably) minimum variance. However, in practice, asset returns are not automatically so scaled.

Suppose, in the same notation as above,
\begin{equation}
    R_{Ai} \sim \sqrt{V_A}\sqrt{\Delta_i}\Nor_{a}
\end{equation}
with $V_A$ the asset's variance. Then an unbiased estimator $\widehat V$ is easily seen to be
\begin{equation}
    \widehat V_{Ai} = R_{Ai}^2 / \Delta_i
\end{equation}
We can readily calculate by the same methods as above,
\begin{equation}
    \Var(\widehat V_{Ai}) =  \frac{\Var(R_{Ai}^2)}{\Delta_i^2} = 2V^2
\end{equation}
The useful result is that the answer does not depend on $R_{Ai}$ or $\Delta_i$, hence not on $i$ at all. This implies that the optimal weighted average of the $\widehat V_{Ai}$'s has constant weights. Thus,
\begin{equation}
    \widehat{\widehat V}_A = \frac{1}{M}\sum_{i=1}^{i=M} R_{Ai}^2 / \Delta_i
\end{equation}

With this result, we can write a self-scaled version of equation \eqref{rhonearopt}, here expanded in full glory with $I$ an index over overlapping pairs of $A$ and $B$ segments, $i$ and $j$ indices over $A$ and $B$ segments individually:
\begin{equation}
    \widehat\rhohat = \frac{\displaystyle \sum_I\frac{1}{C_{II}}\frac{R_{AI}R_{BI}}{\Delta_{A\cap B,I}}}{\displaystyle \left(\sum_I \frac{1}{C_{II}}\right) \sqrt{\frac{1}{M}\sum_i\frac{R_{Ai}^2}{\Delta_{Ai}}}\sqrt{\frac{1}{K}\sum_j\frac{R_{Bj}^2}{\Delta_{Bj}}}}
\label{scaled}
\end{equation}
with $C_{II}$ given by equation \eqref{CII}. Equation \eqref{scaled} is the main result of this paper and is the form that we use in the simulations that now follow.

\FloatBarrier
\section{Compare Optimal and Fast Estimators\\in Simulation}
\label{sec4}

Within a simulated fixed time interval, one easily generates $M$ Poisson random time ticks for asset $A$ (the more frequently traded) and $K \le M$ time ticks for asset $B$. Enumerating the overlapping segment pairs $I$ as discussed in Section \ref{sec23}, we can construct the correlation matrix $\Cbf$, from which can be calculated the variance of our two proposed estimators for $\rho^2$, the optimal one, equation \eqref{rhoopt}---which requires inverting $\Cbf$---and the fast estimator, equation \eqref{rhonearopt}---which is simply the inverse-variance weighted average of the individual variances $\Var(\rhohat_I)$, equation \eqref{eq7b} or \eqref{CII}.

For the purpose of this section, namely comparing the variances of the optimal and fast estimates of $\rho$,  there is no need any simulated returns data ($R_A$ or $R_B$). We need to simulate only the tick times. When $M$ and $K$ are both large, as for large data sets of tickwise trades, the meaningful free parameters in this simulation are $K/M$, the ratio of average trading rates, and $\rho$.

For summarizing the simulation results, the various special cases in Section \ref{sec6} suggest that factor out some expected dependencies and write
\begin{equation}
    \Var\left(\widehat\rhohat\right) = (1 + \rho^2)\left(
    \frac{1}{K} + \frac{1}{M}\right)\,\times\,
    F_{\genfrac{}{}{0pt}{2}{\text{optimal, or}}{\text{fast}}}
\label{eq30}
\end{equation}
where the ``correction factors" $F_{\text{optimal}}$ or $F_{\text{fast}}$, the output of the simulations, should be close to unity. The expression $(1/K + 1/M)$ stands in for $1/\text{min}(K,M)$, embodying the idea that the variance should be dominated by the less-frequently traded asset.

Tables S1 and S2 in the Supplementary Information give
simulation results for $F_{\text{optimal}}$ and $F_{\text{fast}}$
across a range of trading rate ratios $0.02 \le K/M \le 1$ and correlation values $0 \le \rho^2 \le 1$. An adequate summary of these simulation results is that the correction factors $F_{\text{optimal}}$ and $F_{\text{fast}}$ both differ from unity by at most $\sim 20$\%. Using equation \eqref{eq30} with $F=1$ is thus good enough for most purposes, e.g., estimating how many ticks are needed to estimate $\rho$ to a specified approximate accuracy.

Table S3 shows results for the ratio $F_{\text{fast}}/F_{\text{optimal}}$. As expected, the values are all greater than unity, since the fast estimator is not optimal, but all are less than $\approx 1.04$, this only in the extreme case $\rho=1$ and $K=M$. This surprisingly close equality of the two shows that the fast estimator is, in practice, about as good as the optimal estimator.

An important note is that, while the fast estimator $\widehat\rhohat$ is nearly optimal, the naive estimate of its accuracy using the standard formula for inverse-variance weighted averages $[\sum_I (1/C_{II}]^{-1}$, is far off, in fact often misleadingly small. Rather, equation \eqref{eq30}, with or without Table S2's correction, estimates the error accurately.

\section{Compare Fast Estimator to Grid Interpolation}
\label{sec5}

Previous work (e.g., \cite{Jong,RealCorr,Ole,IntroBook,Bonanno,Geo}) typically deals with the problem of asynchronous tick times by some form of interpolation or extension of the total returns of individual assets to a common regular grid. On such a grid, the correlation of two assets, $A$ and $B$, is readily estimated by
\begin{equation}
    \widehat\rhohat = \frac{\sum_i R_{Ai}R_{B_i}}{\sqrt{\sum_i \left(R_{Ai}\right)^2}\sqrt{\sum_i \left(R_{Bi}\right)^2}}
\label{gridest}
\end{equation}
where the index $i$ sums over grid intervals.

First observed by Epps \cite{Epps} on actual trading data, estimates using equation \eqref{gridest} are sensitive to the choice of grid interval $\Delta t$, with smaller intervals yielding smaller apparent correlations $\widehat\rhohat$. For many years puzzling, it is now generally accepted \cite{Zhang} that this ``Epps effect" is largely an artifact of equation \eqref{gridest}'s being a biased statistical estimator. Decreasing $\Delta t$ implies fewer ticks per grid interval, producing an increasing bias. The total error in equation \eqref{gridest} is the sum in quadrature of its statistical error (variance), which decreases with the number of grid intervals and its bias, which decreases with the number of ticks per grid interval.

\begin{figure}[ht]
\centering
\includegraphics[width=14cm]{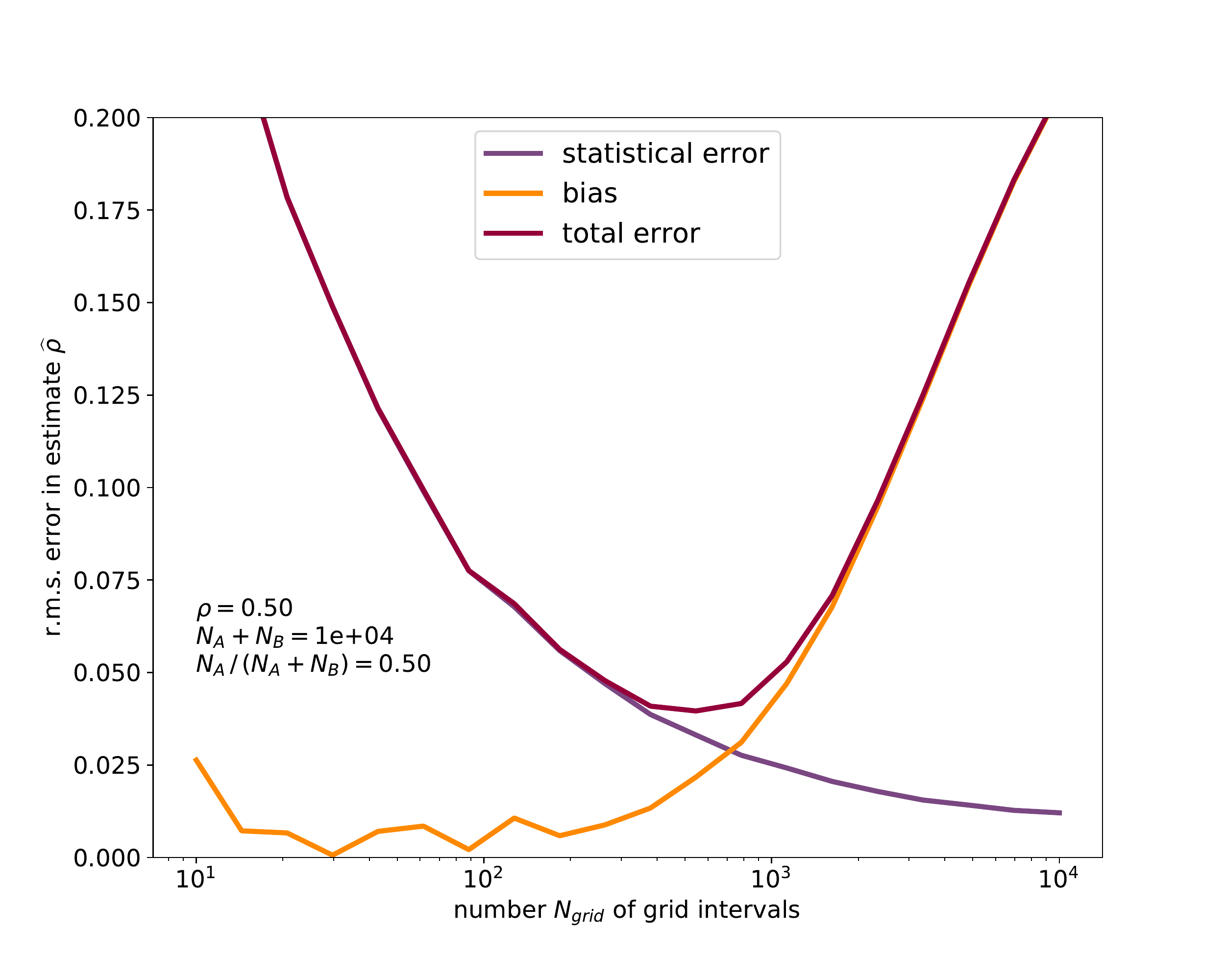}
\caption{The grid interpolation method, not here recommended, evidences a classical bias-variance tradeoff as a function of the grid spacing. Here simulating 5000 ticks each of assets $A$ and $B$ with correlation $\rho=0.5$, the smallest error in the estimate $\rhohat$ is obtained by interpolating onto a grid of size $\approx 600$. But, that value changes with the number of ticks, the ratio of trading rates, and the (unknown) value of $\rho$, not in a simple way. For an inoptimal value, the error in $\rhohat$ is much larger.}
\label{fig1}
\end{figure}

Figure \ref{fig1} illustrates this variance-bias tradeoff. We simulate $N_A = N_B = 5000$ ticks and corresponding returns for two correlated assets, $A$, and $B$, at independent Poisson random times and interpolate the returns onto a grid. The coefficient of correlation is $\rho=0.5$. The independent variable in the figure is the size of the grid interpolated into, ranging from $10$ to $10^4$ grid intervals. A back-of-the-envelope calculation, scaling variance inversely with the number of grid intervals and bias inversely with the number of ticks per interval, suggests that the optimal grid size should scale as a constant times the square root of the number of ticks. In the figure, this is roughly true, however with a constant $\sim 5\times$.

\begin{figure}[ht]
\centering
\includegraphics[width=14cm]{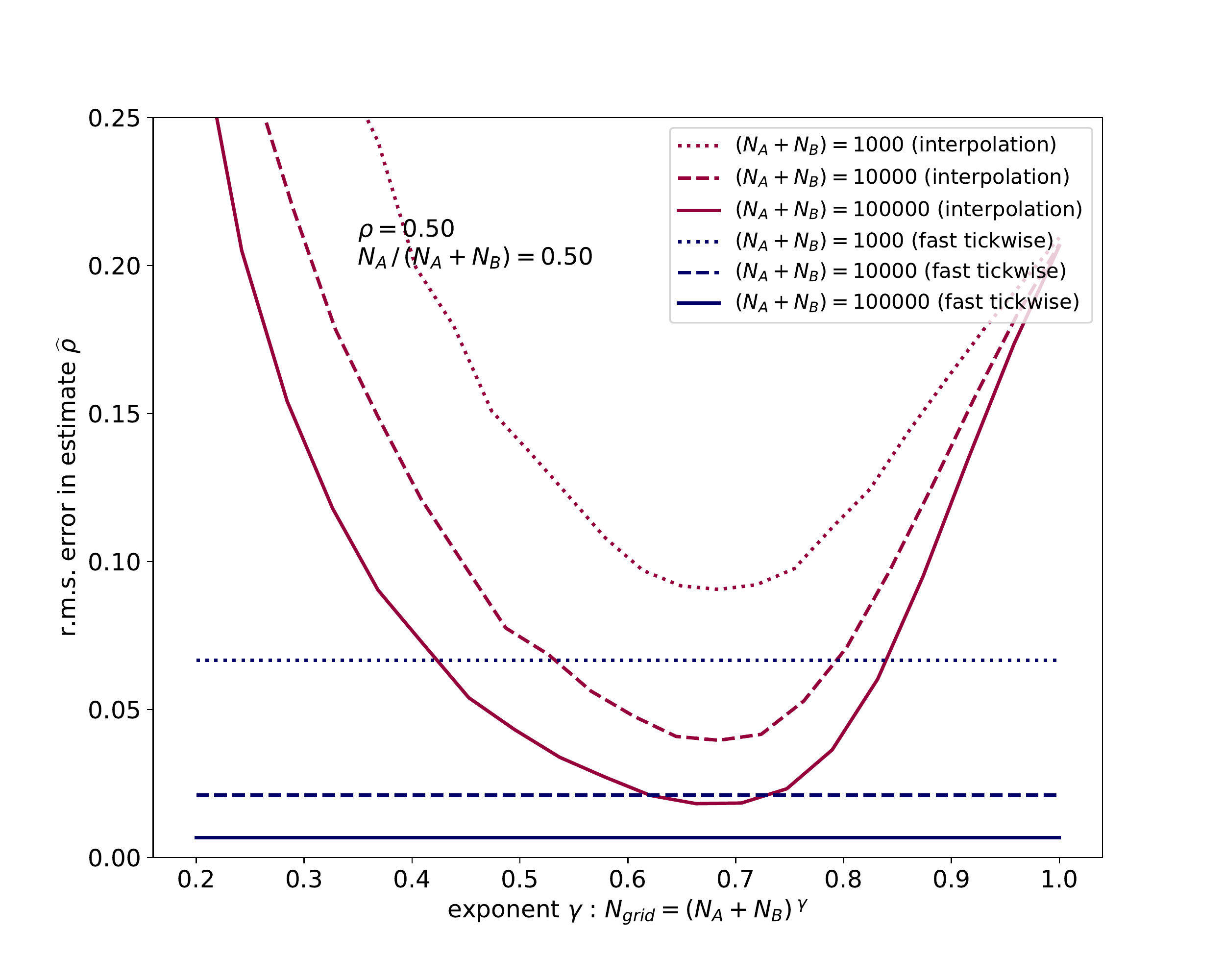}
\caption{Red curves: Like Figure \ref{fig1}, but varying the total number of ticks $N_\text{ticks}$ from $10^3$ (dotted) to $10^5$ (solid). Blue lines: Total errors for the same simulations using the recommended fast tickwise method. (There is no grid, so no dependence on grid size.) Each blue line is everywhere less than its corresponding red curve. Here the abscissa is an exponent relating the grid size to $N_\text{ticks}$, so that (e.g.) $\gamma=0.5$ corresponds to a square root.}
\label{fig2}
\end{figure}

Figure \ref{fig2} tests this scaling law for $N_\text{ticks} = N_A+N_B$ values $10^3$, $10^4$, and $10^5$. It is convenient to let the abscissa be the exponent $\gamma$ in a parameterization
\begin{equation}
    N_\text{grid} = {N_\text{ticks}}^\gamma
\end{equation}
so that (e.g.) $\gamma=0.5$ corresponds to square-root scaling. We see in the figure that the optimal $\gamma$ is approximately constant over the range of $N_\text{ticks}$, but with an apparent value $\sim 0.66$ (or, if actually $0.5$, a substantial and varying prefactor) .

Also shown in Figure \ref{fig2} are the r.m.s. total errors of this paper's fast tickwise estimator, equation \eqref{scaled}. These are horizontal lines since no grid is involved. One sees that, for every simulated value of $N_\text{ticks}$, the fast tickwise estimator yields a smaller error than the interpolation method, even when $N_\text{grid}$ is optimized. When it is not optimized, the fast tickwise estimator is better by often a large factor.

\begin{figure}[ht]
\centering
\includegraphics[width=10.5cm]{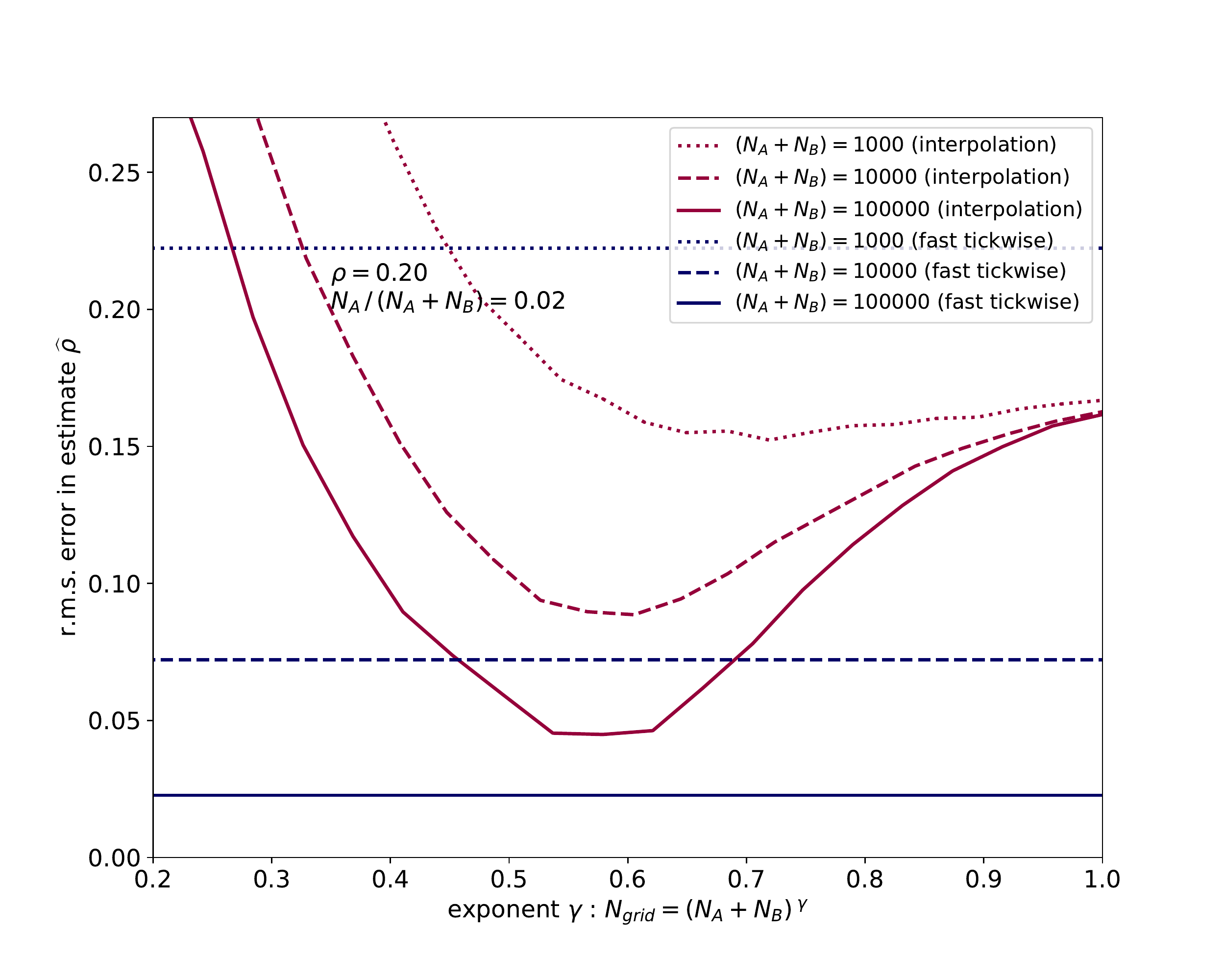}\\
\vspace{-1cm}
\includegraphics[width=10.5cm]{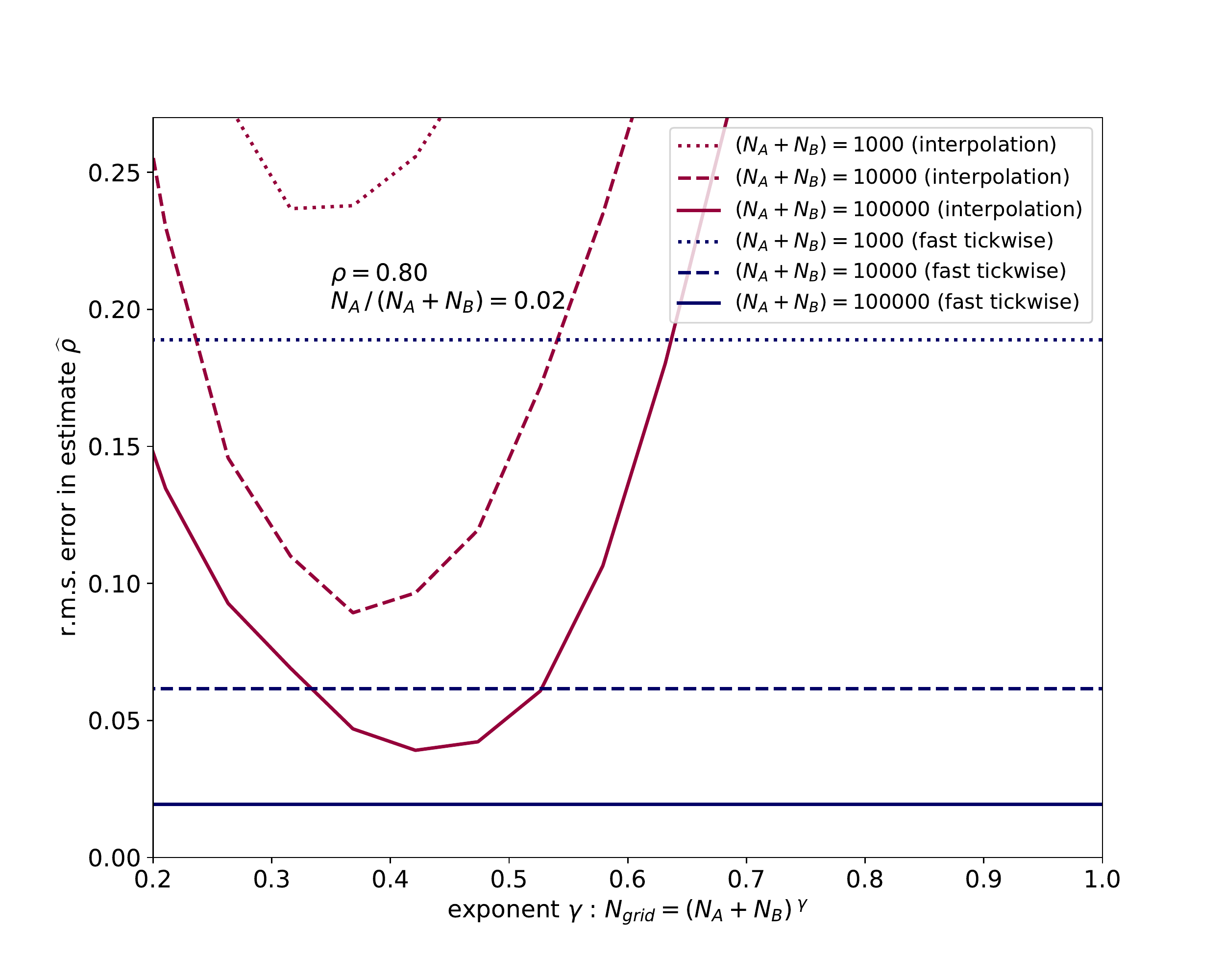}
\caption{Like Figure \ref{fig2}, but with a large ($\approx 50:1$)
ratio of trading rates of the two assets, and for two different values of $\rho$, 0.2 and 0.8. Comparing also with Figure \ref{fig2}, one sees that grid interpolation has no universal optimal exponent $\gamma$ (or universal multiplicative constant if an exponent is assumed). Thus, when $\rho$ is not known a priori, the interpolation method is impractical to optimize. The blue lines, showing the measured error of the fast tickwise method, are everywhere less than their corresponding red curve.}
\label{fig3}
\end{figure}

In fact, optimizing $N_\text{grid}$ for the interpolation method is likely not practical, except in very special cases. In simulations, we find the optimal value of $N_\text{grid}$ to be a function of both $\rho$ (of course not known in advance) and of the tick rate ratio $N_B/N_A$. Figure \ref{fig3} illustrates the complexity for the case $N_B/N_A \approx 0.02$ (1:50 trading ratio) and for $\rho=0.2$ and $\rho=0.8$. The tradeoff curves are seen to be quite different, and not always scaling with any single exponent. Iterative schemes of improving $\rho$ by re-estimation of the optimal $N_\text{grid}$ seem likely to be foiled by the large off-optimum errors spoiling convergence. (This is analogous to the well-understood fools errand of ``improving" an estimator by ``subtracting off its bias", see, e.g., \cite{yudkowsky}.)

Given the superiority of the proposed fast tickwise estimator in simulations with realistic parameters, we have not investigated the pathologies of the interpolation method in more detail. In the limits $\rho\rightarrow 0$ and $\rho\rightarrow 1$, there are cases where the interpolation method's formal total errors are less than the tickwise estimator; but this occurs with respectively $N_\text{grid} \gg N_\text{ticks}$ (most grid intervals have zero ticks, giving an artifactual bias towards zero) and $N_\text{grid}\rightarrow 1$ (all ticks in a single interval, giving large statistical error unless, strictly, $1-\rho \ll 1$). For completeness, the Supplementary Information includes figures analogous to Figures \ref{fig2}--\ref{fig3} for these artifactual cases.

\FloatBarrier

\section{Discussion and Summary}
 The main results of this paper are equation \eqref{eq19}, leading to equations \eqref{CII} and equation \eqref{scaled}, which together yield an estimate $\widehat\rhohat$ (termed ``fast tickwise estimator") of the correlation $\rho$ between two assets from their asynchronous tick-by-tick trading prices.
 
 Equally important is equation \eqref{eq30} (with or without the near-unity correction factors in Table S2), which estimates the variance of the estimator $\widehat\rhohat$. Equation \eqref{eq30} implies that $\rho$ can be determined with an accuracy $\delta\rho$ with roughly $(1+\rho^2)/(\delta\rho)^2$ ticks of the slower-trading asset, or about twice that number for two assets trading at the same average rate.

 The fast tickwise estimator is calculable in time linear in the total number of ticks $N_\text{ticks}$. In simulations with random trading times (\S\ref{sec4}), its estimates $\widehat\rhohat$ differ only negligibly from its related unbiased, minimum-variance estimator, equation \eqref{rhoopt}, whose workload is much greater (between quadratic and cubic in $N_\text{ticks}$, depending on the ratio of the two trading rates).

 In comparison with the common alternative of interpolating the total returns of each asset as a function of time onto a regular grid of times, then estimating their correlation on this grid by standard methods (e.g., equation \eqref{gridest}), the fast tickwise method generally more accurate by a large factor (implying the square of this factor in the number of ticks needed for a specified accuracy). Exceptions are cases where the interpolation grid spacing is fine-tuned for a pre-determined value of $\rho$, not in general useful when $\rho$ is itself the quantity to be estimated. In these cases the fast tickwise method is more accurate, but by a smaller factor (Figures \ref{fig2}--\ref{fig3}).

\FloatBarrier
\printbibliography

@article{LoMac,
  author  = {Andrew W. Lo and A. Craig MacKinlay}, 
  title   = {An Econometric Analysis of Nonsynchronous Trading},
  journal = {Journal of Econometrics},
  year    = 1990,
  pages   = {181-211},
  volume  = 45
}

@article{Jong,
  author  = {Frank de Jong and Theo Nijman}, 
  title   = {High frequency analysis of lead-lag relationships between financial markets},
  journal = {Journal of Empirical Finance},
  year    = 1997,
  pages   = {259-277},
  volume  = 4
}

@article{Bonanno,
  author  = {Giovanni Bonanno and Fabrizio Lillo and Rosario N. Mantegna}, 
  title   = {High-frequency cross-correlation in a set of stocks},
  journal = {Quantitative Finance},
  year    = 2001,
  pages   = {96-104},
  volume  = 1
}

@article{Reno,
  author  = {R. Reno}, 
  title   = {A closer look at the Epps effect},
  journal = {Int. J. Theor. Appl. Finance},
  year    = 2003,
  pages   = {87-102},
  volume  = 6
}

@article{Precup2,
  author  = {Ovidiu V. Precup and Giulia Iori}, 
  title   = {A comparison of high-frequency cross-correlation measures},
  journal = {Physica A},
  year    = 2004,
  pages   = {252-256},
  volume  = 344
}

@article{Ole,
  author  = {Ole E. Barndorff-Nielsen and Neil Shephard}, 
  title   = {Econometric Analysis of Realized Covariation: High Frequency Based Covariance, Regression, and Correlation in Financial Economics},
  journal = {Econometrica},
  year    = 2004,
  pages   = {885-925},
  volume  = 72
}

@techreport{RealCorr,
  title       = {Realized Correlation Tick-by-Tick},
  author      = {Fulvio Corsi and Francesco Audrino},
  institution = {University of St. Gallen Department of Economics},
  address     = "CH-9000 St. Gallen",
  number      = "2007-02",
  year        = 2007,
  month       = jan
}

@article{Precup1,
  author  = {Ovidiu V. Precup and Giulia Iori}, 
  title   = {Cross-correlation Measures in the High-frequency Domain},
  journal = {The European Journal of Finance},
  year    = 2007,
  pages   = {319-331},
  volume  = 13
}

@article{Zhang,
  author  = {Lan Zhang }, 
  title   = {Estimating covariation: Epps effect, microstructure noise},
  journal = {Journal of Econometrics},
  year    = 2011,
  pages   = {33-47},
  volume  = 160
}

@article{Geo,
  author  = {K. Rehfeld and N. Marwan1 and J. Heitzig1 and J. Kurths}, 
  title   = {Comparison of correlation analysis techniques for irregularly sampled time series},
  year    = 2011,
  pages   = {389–404},
  volume  = 18
}

@book{IntroBook,
  author    = {M. Dacorogna and R. Gencay and U.A. Muler and R.B. Olsen and O.V. Pictet},
  title     = {An Introduction to High-Frequency Finance},
  publisher = "Academic Press",
  address   = "New York, NY",
  year      = 2001
}

@article{Epps,
  author  = {Thomas W. Epps}, 
  title   = {Comovements in Stock Prices in the Very Short Run},
  journal = {Journal of the American Statistical Association},
  year    = 1979,
  pages   = {291-298},
  volume  = 74
}

@misc{yudkowsky,   
    title = {Useful Statistical Biases},   
    url = {https://www.lesswrong.com/posts/Wwq6WFpx9HyzwgCKx/useful-statistical-biases},   
    author = {Eliezer Yudkowsky},   
    year = {2007},   
    note = {Accessed on February 11, 2023} 
}

\newgeometry{top=4cm}
\section*{Supplementary Information}
\begin{table}[ht]
\small
\centering
\begin{tabular}{ l | r c c c c c c c } 
\hline
$\rho^2$  & $K/M=$ 1.00 & 0.50 & 0.25 & 0.11 & 0.05 & 0.04 & 0.03 & 0.02 \\
\hline
0.00 & 1.220 & 1.204 & 1.165 & 1.110 & 1.065 & 1.051 & 1.036 & 1.014 \\
0.25 & 1.174 & 1.162 & 1.129 & 1.087 & 1.048 & 1.039 & 1.024 & 1.005 \\
0.50 & 1.141 & 1.130 & 1.104 & 1.068 & 1.038 & 1.029 & 1.018 & 1.001 \\
0.75 & 1.114 & 1.106 & 1.086 & 1.055 & 1.030 & 1.021 & 1.011 & 0.997 \\
1.00 & 1.094 & 1.087 & 1.069 & 1.045 & 1.023 & 1.015 & 1.007 & 0.993 \\
\hline
\end{tabular}
\caption{Factor $F_\text{optimal}$ in equation \eqref{eq30}, giving the variance of the minimum variance (but computationally expensive) estimator of the correlation $\rho$}
\vspace{0.5cm}
\begin{tabular}{ l | r c c c c c c c } 
\hline
$\rho^2$  & $K/M=$ 1.00 & 0.50 & 0.25 & 0.11 & 0.05 & 0.04 & 0.03 & 0.02 \\
\hline
0.00 & 1.220 & 1.204 & 1.165 & 1.110 & 1.065 & 1.051 & 1.036 & 1.014 \\
0.25 & 1.178 & 1.166 & 1.133 & 1.090 & 1.051 & 1.042 & 1.026 & 1.007 \\
0.50 & 1.155 & 1.144 & 1.118 & 1.080 & 1.047 & 1.037 & 1.024 & 1.006 \\
0.75 & 1.139 & 1.131 & 1.111 & 1.076 & 1.046 & 1.035 & 1.022 & 1.006 \\
1.00 & 1.130 & 1.124 & 1.105 & 1.076 & 1.045 & 1.036 & 1.024 & 1.006 \\
\hline
\end{tabular}
\caption{Factor $F_\text{fast}$ in equation \eqref{eq30}, giving the
variance of the diagonal, inverse-variance weighted (and computationally cheap) estimator of the correlation $\rho$}
\vspace{0.5cm}
\begin{tabular}{ l | r c c c c c c c } 
\hline
$\rho^2$  & $K/M=$ 1.00 & 0.50 & 0.25 & 0.11 & 0.05 & 0.04 & 0.03 & 0.02 \\
\hline
0.00 & 1.000 & 1.000 & 1.000 & 1.000 & 1.000 & 1.000 & 1.000 & 1.000 \\
0.25 & 1.004 & 1.004 & 1.004 & 1.003 & 1.003 & 1.002 & 1.002 & 1.002 \\
0.50 & 1.012 & 1.012 & 1.012 & 1.011 & 1.008 & 1.007 & 1.006 & 1.005 \\
0.75 & 1.022 & 1.023 & 1.023 & 1.020 & 1.015 & 1.013 & 1.011 & 1.009 \\
1.00 & 1.033 & 1.034 & 1.034 & 1.030 & 1.022 & 1.020 & 1.017 & 1.013 \\
\hline
\end{tabular}
\caption{Ratio of Table 2 to Table 1, that is, $F_\text{fast}/F_\text{optimal}$ in equation \eqref{eq30}. That all values are close to unity demonstrates that the fast tickwise estimate of the correlation $\rho$ is, for practical purposes, optimal.}
\end{table}
\restoregeometry
\newgeometry{top=1cm}
\begin{figure}[ht]
\centering
\includegraphics[width=12cm]{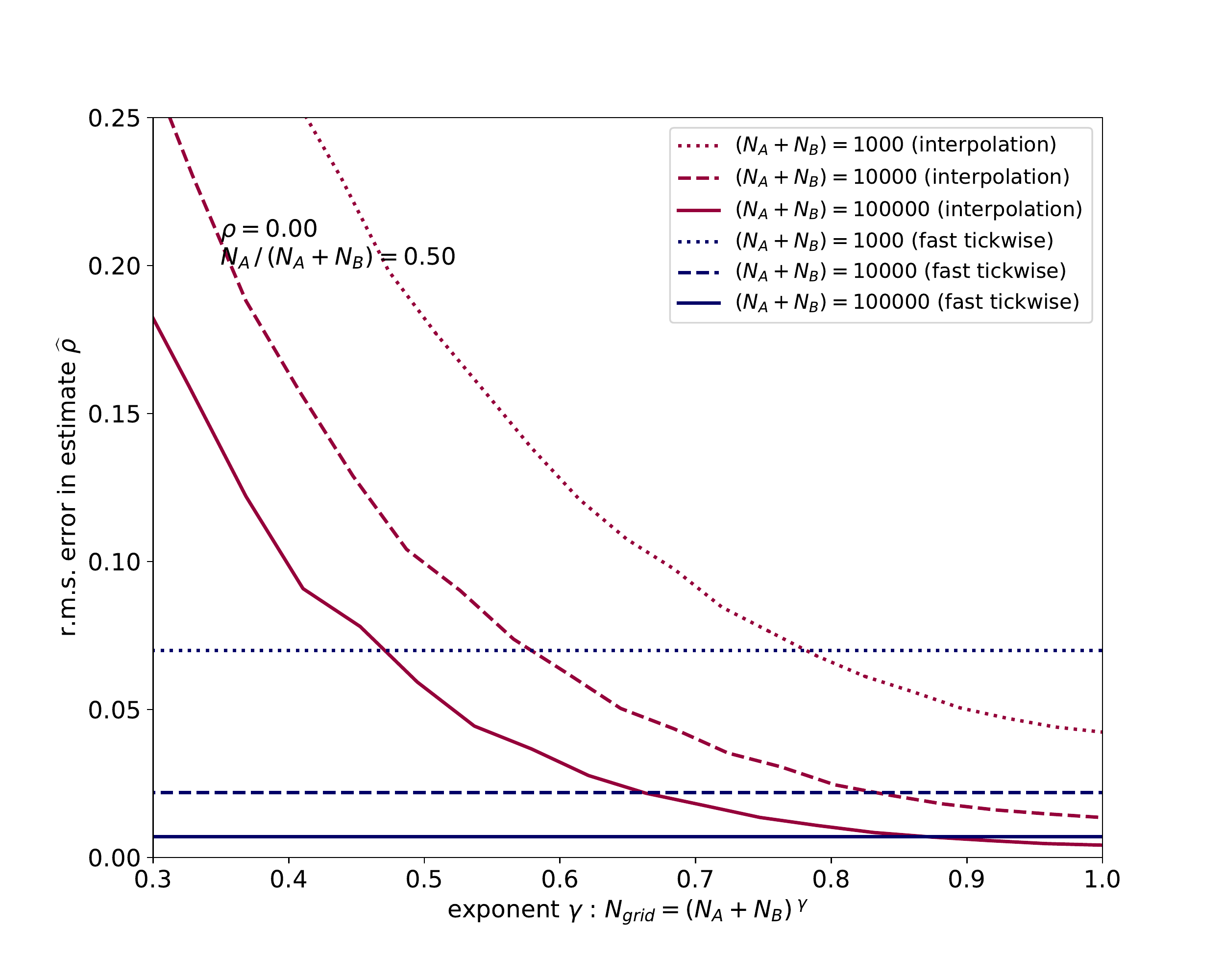}
\caption{Like Figures \ref{fig2} and \ref{fig3}, but extreme case of $\rho=0$. Here the interpolation method has smaller formal errors than the fast tickwise method when the grid size $N_\text{grid}$
is $\gtrsim N_\text{ticks}$. But, in this limit, the bias towards zero is large for any actual value of $\rho$, so the accuracy is an artifact of bias towards what happens to be the correct value.}
\includegraphics[width=12cm]{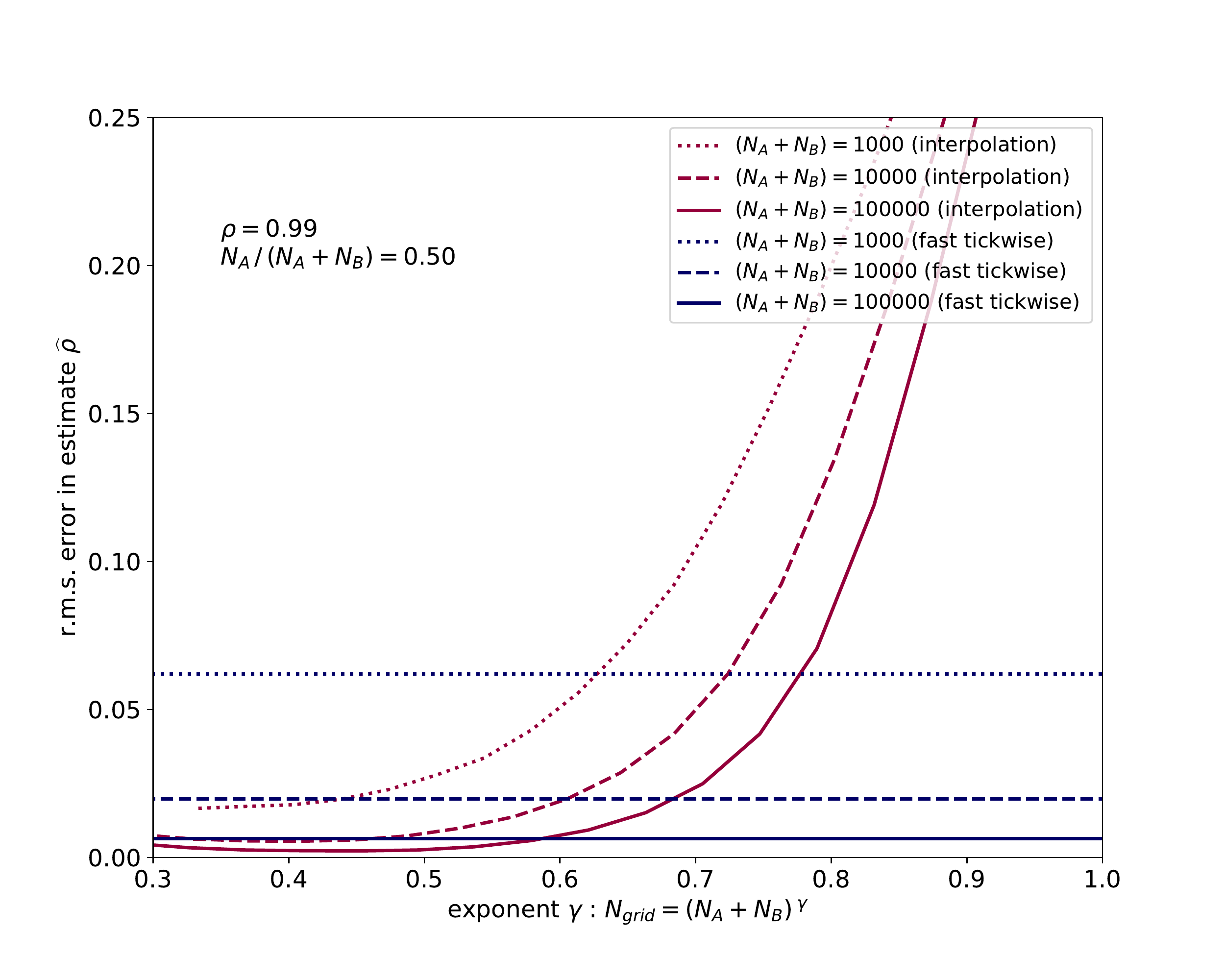}
\caption{Same as above, but with $\rho=0.99$. Here the optimal interpolation grid has only a single interval. The small error is an artifact of $1-\rho \ll 1$. For any larger difference, the statistical error of the single-interval estimate would be huge (compare Figure \ref{fig1}.) }
\end{figure}
\restoregeometry
\end{document}